\documentclass[11pt,a4paper]{article}
\usepackage{amsmath,amssymb,graphicx,float,latexsym,psfrag,subfigure}
\usepackage{color}
\usepackage{overpic}

\begin{document}
\title{The prospect of using LES and DES in engineering design, and the research required to get there}
\author{Johan Larsson\thanks{Department of Mechanical Engineering,
    University of Maryland. E-mail: jola@umd.edu.}~
  and
  Qiqi Wang\thanks{Department of Aeronautics and Astronautics,
    Massachusetts Institute of Technology. E-mail: qiqi@mit.edu.}}
\maketitle

\begin{abstract}
In this paper we try to look into the future to envision how large eddy
and detached eddy simulations (LES and DES, respectively) will be used
in the engineering design process about 20-30 years from now.
Some key challenges specific to the engineering design process are
identified, and some of the 
critical outstanding problems and promising research directions
are discussed.
\end{abstract}

\section{Introduction}

This paper is about the feasibility of using
large eddy simulations (LES) and (to a somewhat lesser extent) detached
eddy simulations (DES)
in the industrial engineering design process about 20-30
years from now.
Many practitioners of  
computational fluid dynamics (CFD) for realistic turbulent flows
believe
that the cost
of LES is and will remain so high that it
won't truly enter the practical engineering design optimization
process for another 30 years.
Additionally, although anybody can create a grid and run an LES these
days, detailed expertise is required to ensure that the result are
trustworthy and to interpret the results.
The verification process is either lacking or completely
expert-driven, and requires large amounts of both human and computer time.

Prior to writing this paper, we communicated with CFD experts at
different corporations in
a range of business areas;
one message was that,
despite the difficulties, there is a true need for LES in design optimization.
LES
combines the advantages of simulations with (some of) the
reliability of physical experiments.
It gives engineers higher
confidence in the results,
especially in the complex flows typically occuring in optimized engineering devices
(e.g., jet engine combustors)
or near the edges of the operational envelope (e.g., an airfoil near stall),
than the lower fidelity simulations primarily used today (i.e., RANS
and URANS).
With more accurate predictions for complex and off-design flows,
engineers could push designs closer to the edge of the envelope with lower safety
factors and possibly more optimal designs (faster, cheaper, greener, etc).

In this paper we try to  look into the future to envision how
LES will be used in the engineering design process about 20-30
years from now.
The context of the industrial design process and the challenges this
creates is discussed in section~\ref{sec:challenges}; the research
required to get there is
discussed in section~\ref{sec:research}.
While our focus is primarily on LES, much of the discussion applies
to DES as well.

At the outset, we want to be emphasize that this paper is not a review
paper.  We do not intend to cover all the available literature.
As requested by the organizer of this special issue,
we instead attempt to be ``prophetic'' -- looking into the future to explore
what we see as the key challenges of using LES in engineering
design, and the research required to get to where we think LES should be
in 20-30 years.

\section{The challenges of using LES in engineering design \label{sec:challenges}}

LES is a well-established and very commonly used tool within the
academic community.
It is certainly used by engineers in industry, but rarely within the
actual design process; more often, LES is used in one-off situations
or in industrial R\&D studies.

It is crucial to appreciate
just how different a design process is from an academic study.
The goal of the latter is often to generate physical insight, and thus
the fidelity and accuracy of the LES are very important.
The typical academic study runs for 3-5 years, during which perhaps a
handful of high-quality LES are performed.

The engineering design process is very different.
The objective is to make the correct design decision, and later to
certify this.
The quality of any given simulation performed to support this design
decision is of less importance, provided that it leads to the correct
decision.
The overall time-scale of the design process, from concept to finished
product, may not be that different from academic studies, but the
time-scale of decision-making on sub-systems certainly is.

Taking a big-picture view of the design process, the simulation-needs
are clearly different at different stages.
Choosing between different concepts may require very fast,
low-fidelity tools.
As the design converges towards an optimized final product,
higher-fidelity tools may be required for the final optimization.
LES combines the advantages of simulations with some of the
trustworthiness of experiments; it clearly has a place somewhere in
the design/optimization process.
The question is where and how to integrate it with other tools.

\subsection{The need for fast turn-around}

\begin{quote}
I have not failed. I've just found 10,000 ways that won't work --
Thomas A. Edison
\end{quote}

Engineers work fast, because
they often evaluate many designs before finding one that works.
They need to quickly evaluate new designs, starting from
a CAD model, to a CFD mesh, to a flow solution, to the post-processed
result.
The relative maturity of many engineering fields contributes to this
fact: we already know how to design things that are close to optimal,
and the last few percent of performance are the most difficult to attain.
This implies that many design candidates need to be assessed, but also that a
relatively high degree of accuracy may be needed (to distinguish
between those last few percent).
The latter is a saving grace for LES, as it implies that it can be
useful for design well before it becomes an over-night tool like RANS
-- as long as the added-value of an LES study is sufficiently
greater than a RANS study,
a longer turn-around time can be acceptable.
Turn-around times of a few
days are probably acceptable for many applications.

The speed that matters is that of the full simulation process, from
CAD to grid to solution to verification to distilled results.
All of these steps are currently bottlenecks, and all must be
substantially improved in order to enable fast turn-arounds.
Automation may become as important as speed: the cost of computer-time
is ever decreasing, while the cost of (highly educated) human-time is
ever increasing.

\subsection{The need for more than a single prediction}

Engineers have labored the iterative process of design optimization
since the invention of the wheel.  To engineer the best product,
we design, assess, then keep redesigning and reassessing.
For LES and DES to benefit engineering design, these techniques must be able to not
only assess a single design, but help us make decisions as
we iteratively redesign and reassess.

Critical to making these decisions is how the flow field
depends on the design.  We want to know
how quantities of interest, which collectively quantify how good the
design is, depend on the design variables that parameterize all
possible designs.  Knowing this dependence enables an engineer to focus
more on designs that produce favorable flows while
avoiding designs that induce undesirable ones.  To understand this
dependence requires more than a single prediction by a single
simulation.  It requires integrating LES with the tools of
sensitivity analysis.

Particularly useful for efficient design optimization is local
sensitivity analysis, which computes the derivatives
of the quantities of interest with respect to the design variables.
The derivatives are essential in driving
gradient-based optimization algorithms.  They enable these algorithms
to improve a design more quickly than gradient-free ones, especially
when there are many design variables.

Sensitivity analysis is also useful in uncertainty quantification (UQ), a
key component of engineering design.  It helps engineers measure how
uncertain factors affect the performance of their design.  These
factors include manufacturing variability, operating conditions,
aging and degradation.  If they cause
unacceptable effects in performance, engineers need to know
so that they can take measures to
control the source of uncertainty, or to make their design more robust.

Despite being useful, sensitivity analysis is problematic for LES and
DES.
As an example, one of the authors once spent more than a year
developing a local sensitivity analysis
code for LES in complex geometries, only to find it diverging
even in the simplest of cases.  He later discovered the cause of
the problem: LES is a chaotic dynamical system, and its
sensitivity diverges due to the ``butterfly effect'' of chaos.
As further explained in Section 3.5,
the divergent sensitivity prohibits engineers from using
this efficient tool in design optimization and in
uncertainty quantification.  Overcoming this divergence and enabling
efficient sensitivity analysis is an important challenge
in making LES and DES suitable for engineering design.

\subsection{Multi-fidelity parametric design}

Iterative design offers not only challenges, but also an
opportunity for LES and DES to guide how reduced-order models are used.
Scientists and engineers have developed models of many levels of
fidelity for flow fields.  They range from models derived from
physical insight, often represented as algebraic or ordinary
differential equations, to potential flow models that solves
steady and linear partial differential equation, to RANS which solves
a set of steady but nonlinear partial differential equations.
LES (and, to a lesser extent, DES)
tops almost all models both in complexity and
in fidelity.  LES performed for one particular design reveals much
flow physics, which should guide engineers in choosing a reduced-order
model, and even in tuning these models to make them more accurate for
similar designs.

LES and DES provide new opportunities in optimization methods.
They force us to think of optimization beyond its classic
definition, of minimizing an objective function that can be
evaluated exactly at fixed cost.  In a multi-fidelity design,
different models evaluate the objective function with different
accuracy.  High fidelity models, like LES and DES, have additional
sampling errors unless they run for infinite time.  Optimization becomes a
resource management problem of how to allocate computation time between
low and high fidelity models, such that they jointly minimize
an objective function that can be evaluated at various levels of
fidelity at various costs.

In this view, the development of a large aircraft, which takes many
years, can be viewed as a single optimization process.  In its early
stage, a wide range of designs are explored using low fidelity estimates
of the objective functions and constraints.  As the optimization
converges, it uses higher fidelity tools, upgrading from simulations to
wind tunnel models and ultimately to test flying full scale aircrafts.
LES combines the efficiency of simulation and trust-worthiness of
physical experiments.  It surely has its place in optimization.  The
question is where it fits best, how it would first be integrated into
the process, and how it would make the design process faster and the
product better.

Optimization is very expensive, so it would be completely unreasonable to think that we
will do optimization solely with LES or even DES in 20 years -- but the challenge
is instead to think of ways to do optimization where we can utilize
the strengths of LES.  Doing so collaboratively with other models is
both a research opportunity and a challenge.

\subsection{The next generation computer hardware \label{sec:hardware}}

High performance computing gets 1000 times faster every 10 to 15
years.  This trend has lasted for several decades, and is
predicted to continue for the next 20 years.
Studies at the US Department of Energy have proposed extreme numbers of energy-efficient
cores running in parallel to achieve ``exascale'' computing
\cite{kogge2008exascale,amarasinghe2009exascale,springerlink10}.
This trend will accelerate future LES calculations with
thousand-fold faster floating point operations, but will also challenge it
with a high cost of communication and data movement and extreme penalties on
serial algorithms.

One impact on LES is that it will likely make spatially implicit
differencing and filtering schemes~\cite{lele:92} less attractive.
These schemes, like Fourier transforms, introduce global
connectivities, which will presumably be barriers to exascale
parallelism.

Another impact is that we will likely need intricate computer science
in future CFD codes.
When the 17 Pflop/s, 1.6M core, BlueGene/Q \emph{Sequoia}
supercomputer was deployed in 2012-2013, a few LES-type codes were
tested on it.
The code written by one of the present authors was run on \emph{Sequoia} without
modifications, and utilized about 6\% of the peak
flop-rate~\cite{bermejomoreno:13:supercomputing};
compare this to the very similar code by
Rossinelli \emph{et al.}~\cite{rossinelli:13}
(winner of the 2013 Gordon Bell prize)
which reached 55\% of peak on the same machine with a similar
algorithm (5th order WENO).
To achieve this performance, they had to re-write their code with a
new data layout to maximize memory locality.
If such code- and platform-specific modifications become broadly
necessary on exascale machines, then this will have a large negative
impact on the ability of small research groups to write efficient code
for large-scale simulations.

\section{The research required to get there \label{sec:research}}

There are clearly gaps between the current state-of-the-art in LES and
the requirements outlined in the previous section.
The exponential increase in computer power will help close these gaps,
but clearly will not, by itself, be sufficient.
Research that advances our LES technology is certainly required.

The following sections describe some research areas that we deem
particularly important, promising and exciting.
The topics vary in maturity: some have been part of LES for
a long time, but with important unsolved problems; others are
more novel research areas that we feel are particularly interesting.

\subsection{Complex geometries}

Real engineering systems have complex geometries.
Therefore, it seems clear that any LES developments
that are not applicable to complex and general geometries or to
unstructured grids will not have much impact.
Likewise, developments that cannot be efficiently implemented on
massively parallel computers will likely be of limited value.

These may seem trivial statements, but significant chunks of our
current state-of-the-art in LES does not satisfy them.
For example, the dynamic procedure by
Germano \emph{et al.}~\cite{germano:91}
requires both filtering and averaging in order to work:
the standard academic approach is to perform these steps in
homogeneous directions, which of course does not work in general
geometries.
The Lagrangian averaging approach~\cite{meneveau:96} solves the latter
problem but not the former.
While filtering in non-homogeneous directions is theoretically well
defined, in practice it can lead to a loss in accuracy.
For example, a Germano procedure with wall-normal filtering in a coarsely
resolved laminar boundary layer would return a non-zero eddy
viscosity
--
this error is generally not seen in academic studies where filtering
is performed in homogenous directions only.

A related example is the group of methods that uses Germano's dynamic
procedure in a global sense, to average over the full domain to find a
single, global model coefficient.
This appears to work for academic test cases where a single type of
flow occupies most of the domain, but seems questionable in realistic,
complex situations with many different types of flow within the domain.

\subsection{Wall-modeling \label{sec:wmles}}
The computational cost of LES is essentially independent of the
Reynolds number $Re$ for free shear flows, it is only weakly dependent on
$Re$ for the outer portion (say, the outermost 90\%) of
turbulent boundary layers, but it becomes strongly dependent on $Re$
for the innermost layer (the viscous sublayer, the
buffer layer and the
bottom of the logarithmic layer).
In fact, traditional LES, where the innermost layer is resolved,
permits a grid that is maybe 4 times coarser than DNS in each spatial
direction when applied to high-$Re$ boundary layers, thus a
cost-saving of two orders of magnitude; while significant, this is
hardly a
game-changer, and to truly enable LES of high-$Re$ boundary layer
flows one must model the innermost layer: this is known as ``wall-modeling''.
In a classic paper, Spalart \emph{et al.}~\cite{spalart:97} estimated
that about $10^{11}$ grid points would be needed for wall-modeled LES
of a complete, swept wing at full-scale flight conditions.
While intimidating, this is at least five orders of magnitude
less than would be required for traditional LES \cite{spalart:97}.
The DES approach is a few orders of magnitude cheaper, but by modeling
the full boundary layer it is susceptible to similar errors as
traditional RANS models
in non-equilibrium flows, particularly
(and very importantly)
separating boundary layers~\cite{cfd2030}.
By modeling only the innermost part of the boundary layer, which is
also the part expected to display the closest to universal behavior,
wall-modeled LES is, in principle, capable of more accurate
predictions.

For quite some time, the main outstanding problems in wall-modeling
for LES have been (in the perhaps biased views of the authors):
1) the ``log-layer mismatch'', in which errors near the wall lead to
errors in the predicted skin friction by about 10\%;
2) the inherent assumption of turbulent flow in the wall-models, which
renders them inaccurate in laminar regions and incapable of
predicting transition to turbulence;
and
3) the assumption of equilibrium flow in the wall-models, which
renders their accuracy in non-equilibrium flows (most notably, during
flow separation) questionable.
All of these issues are clearly critical for airfoil flows, and
recent research has made progress on particularly the first two issues.
The log-layer mismatch can be removed by decoupling the thickness of
the modeled near-wall layer from the LES grid, 
thus producing very
accurate predictions of skin friction~\cite{kawai:12:wallmodeling}.
This naturally impacts the accuracy of the predicted frictional drag of an
airfoil,
but also (and arguably more importantly) the predicted separation location at high
angles-of-attack, by ensuring that the incoming boundary layer has the
right near-wall momentum.

\begin{figure}[t]
\centering
\includegraphics[width=120mm]{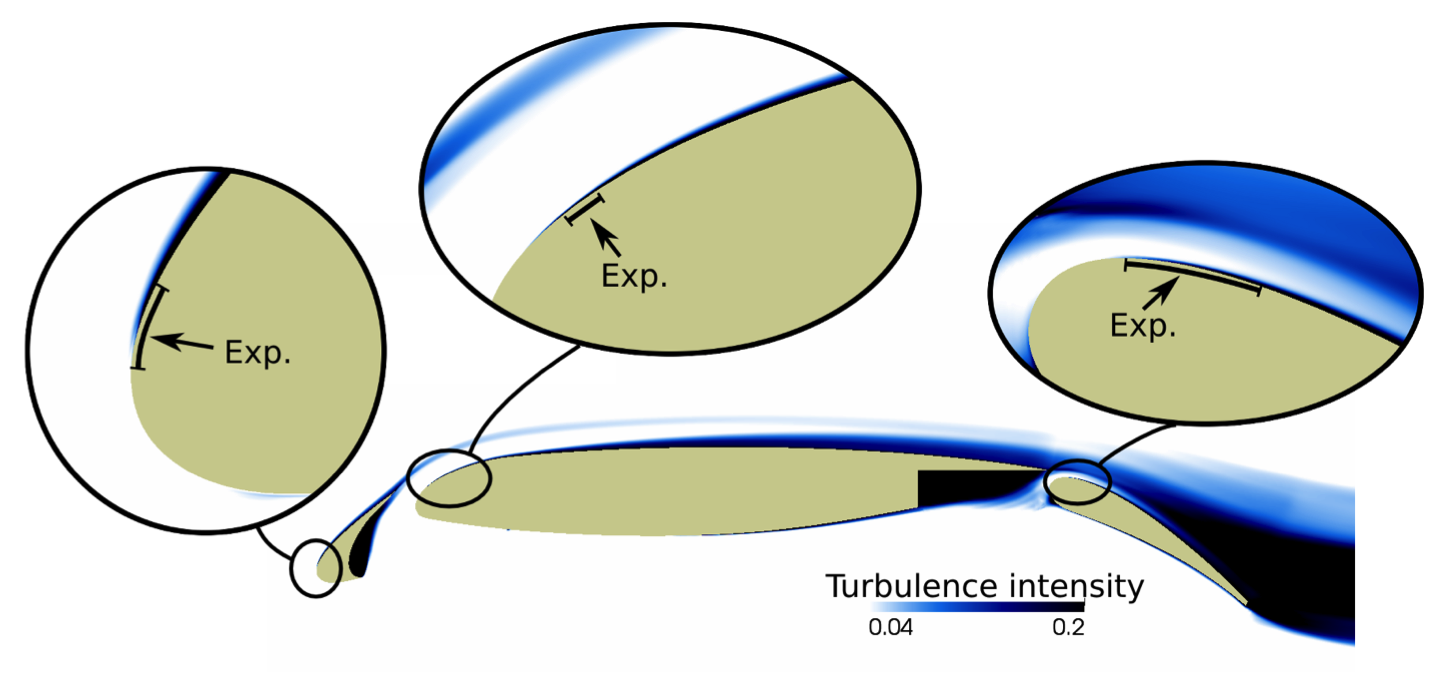}
\caption{ \label{fig:md30p30n}
  Predicting laminar-to-turbulent transition using wall-modeled LES on
  the MD 30P/30N airfoil with deployed slat and flap at $19^\circ$
  angle-of-attack and chord Reynolds number of $Re_c = 9\cdot
  10^6$.
  The predicted transition occurs in the location, on each airfoil
  component, where the turbulence intensity rapidly increases to a
  finite value.
  The experimentally measured transition locations are indicated as
  ranges.
  Figure adopted from~\cite{bodart:12}.
}
\end{figure}

In more recent work, Bodart \& Larsson~\cite{bodart:12} introduced a
sensor into the wall-model, which utilizes information from the
resolved LES region to decide whether the boundary layer is laminar or
turbulent, and only applies the wall-model if the latter.
This was shown to enable the prediction of a transitional turbulent
boundary layer using wall-modeled LES, where the transition location
was in very close agreement with DNS without being prescribed in any
way.
To show how this can be applied to realistic aerospace problems, 
a wall-modeled LES prediction of the flow around the MD 30P/30N multi-component airfoil
at a realistic full-scale Reynolds number and $19^\circ$
angle-of-attack
is
shown in Fig.~\ref{fig:md30p30n} (taken from~\cite{bodart:12}).
The transition locations were truly predicted by the wall-modeled LES
method, without being prescribed in any way.

The question of how the assumption of equilibrium flow
(basically: attached essentially parallel flow with small pressure
gradient)
in the wall-model affects the accuracy on strongly non-equilibrium
problems
(most importantly, separation)
is less settled then one might imagine.
While a separating boundary layer clearly is not in equilibrium,
the wall-model is actually only applied within the innermost 10\% or
so of the boundary layer.
Provided that most of the non-equilibrium effects are felt primarily
in the outermost 90\%, LES using an equilibrium wall-model might
indeed be capable of very accurate predictions.
Some very recent applications of wall-modeled LES to
shock/boundary-layer interactions support this
view~\cite{bermejomoreno:14:duct},
but much more research is needed to truly settle this matter.
The focus should be on
assessing the predictive ability of
wall-modeled LES on separating and
re-attaching boundary layers, and then, only if necessary, to develop more
complex models.

\subsection{Time integration}
For the steady and unsteady RANS simulations used predominantly in
industry today, fully implicit solvers are the most efficient.
A fully implicit solver is more
expensive (per time step) than a fully explicit method, but the latter is, of course,
only stable for time steps smaller than some
$\Delta t_{\rm stab}$.
The exact cost difference between explicit and implicit methods
depends strongly on the problem at hand, but for compressible viscous
flows the difference can be many orders of magnitude.
Simplistically, the fully explicit method is the more efficient one if
the time step required for
stability $\Delta t_{\rm stab}$ is no more than
about 2-3 orders of magnitude
smaller than
the time step required for
accuracy $\Delta t_{\rm accu}$.

Many academic test cases have simple geometries for which this is true
(e.g., canonical free shear flows);
for these problems a fully explicit method is a good choice.
In other canonical problems this is violated by a linear term in a
single direction
(commonly, the wall-normal viscous diffusion in boundary layer
flows);
for such problems a linear line-implicit treatment is near-optimal in
terms of computational efficiency.
In many real-world problems, however, the time steps
$\Delta t_{\rm stab}$ and $\Delta t_{\rm accu}$ differ by many, many orders
of magnitude without there being any simple way to alleviate the
problem;
one thus faces a devil's choice between taking millions of time steps
using a fully explicit method, or living with the cost (operations and
memory) of solving a fully implicit nonlinear system at every time
step.

The wall-modeled LES of the MD 30P/30N discussed in section~\ref{sec:wmles}
and shown in Fig.~\ref{fig:md30p30n}
can serve as an example of this problem.
Assume a chord of 10 m, which is typical of passenger jets.
The laminar boundary layer near the nose of the slat
is extremely thin due to the strong acceleration in that region,
about 1 mm for our example;
grid cells with a thickness of at most maybe 0.05 mm are required
there to capture the transition process.
The strong acceleration causes the flow to become near sonic above the
slat, and thus a compressible solver is necessary;
the stability limit for an explicit method is set by the
wall-normal grid spacing divided by the speed-of-sound on the slat,
about $\Delta t_{\rm stab} \approx$ 0.2 $\mu$s for our example
(note that the viscous stability limit is much less
severe for this example, by about a factor of $0.25 \Delta y^+ c/u_\tau$).
To compute a single chord-passage might require 0.5M time steps, and
many chord-passages are needed to wash out the initial transient.

To overcome this type of extreme numerical stiffness, a combination of
numerical analysis and physical insight is needed.
The nature of the stiffness is strongly related to physics and how
different physics need to be resolved; thus a purely mathematical
solution will likely not be optimal.
Possible solutions include some form of
selective implicit time stepping algorithm, where only some terms in
some regions of the domain are treated implicitly.
Done right, this should significantly decrease the cost of the
implicit solver.
Alternatively, the problem could be solved through local explicit
time-stepping, with synchronization at regular intervals.
One such algorithm for LES was presented recently by
Lu \emph{et al.}~\cite{lu:14}.
Whatever the approach, in addition to physics-induced stiffness (like
for the MD 30P/30N airfoil), the time-stepping algorithm must also be
able to handle stiffness due to poor grids: when meshing a complex
geometry, it is difficult or impossible to avoid a handful of ``bad
cells'' that might induce tremendous stiffness.

The potential speed-up of improved time-integration methods
is bounded by how much more costly a fully
implicit method is compared to an explicit one; for many LES problems,
a speed-up of about two orders of magnitude should
be possible.

\subsection{Turbulent inflow and initial conditions}

In many applications, one would like to perform LES in only a small
subset of the total domain. Consider a turbofan jet engine: the flow through the
inlet and the compressor is likely sufficiently well captured by RANS,
but the turbulent reacting flow in the combustor would benefit from
being treated by LES~\cite{schluter:04,medic:06}.
There are many other examples.

To perform LES in only a limited part of the domain, one key difficulty
is the boundary conditions, most notably the inflow condition.
Being specified from a lower-fidelity method (RANS or other), the
fluctuating flow will necessarily be approximate in nature.
LES, however, is not accurate until the resolved turbulence is
realistic, with realistic amplitudes, phase relationships and
spectrum;
this occurs after an adjustment region of length
$L_{\rm adj}$,
the value of which depends
on the fidelity and realism of the inflow turbulence~\cite{batten:04,keating:04,xu:04,dimare:06}.
With poor inflows, $L_{\rm adj}$ may be longer than the region one
wants to study; this is clearly a bottleneck holding up the use of LES
on sub-systems within larger, complex systems.

For canonical problems, the recycling/rescaling method of Lund
\emph{et al.}~\cite{lund:98} and its many later improved versions has
enjoyed significant success.
By utilizing the resolved turbulence within the LES itself coupled
with known scaling relationships, very realistic turbulence can be
specified at the inlet.
For complex engineering problems, however, we still need something
better, since the rescaling relationships are then no longer valid.

The inflow problem is related to the problem of requiring long
computation times just to get rid of the initial transients $T_{\rm init}$.
A minimum required $T_{\rm init}$ is for the turbulence to pass
through the domain once; this is proportional to the length of the
domain, and thus a poor inflow may actually increase the cost in a
quadratic manner (more grid points and longer $T_{\rm init}$).
More generally, some flows require realistic initial conditions as
well, with the required $T_{\rm init}$ strongly dependent on the
fidelity of the initial turbulence (channel and pipe flows come to
mind).

Finally, for compressible turbulence, the initial and inflow
conditions will generate acoustic and entropy waves in addition to the
hydrodynamic turbulence.
Depending on the exact method, the level of acoustics may be one or
two orders of magnitude larger than in ``quiet''
turbulence~\cite{ristorcelli:97,larsson:09:inflow}.
In some applications (airframe noise, high-speed transition, \ldots) this is a real problem.

Progress on any of these fronts would give us a greater ability to
aggressively apply LES (or DES) only to the sub-systems or small
portions of the domain where LES is truly needed.
With current inflow technologies, whatever accuracy is gained by the
sub-domain LES may be lost due to the errors incurred at the
RANS $\rightarrow$ LES interface.

\subsection{Grid-adaptation}

In complex geometries, the grid generation process may become very
time-consuming, particularly if high-quality grids that both minimize
numerical errors (smooth grid transitions, low element skewness, etc)
and provide sufficient resolution are needed -- which they are, since
we need trustworthy results for design decision-making.
Adaptive grid-refinement, an iterative process in which the solution
guides how the grid is adapted through some algorithm (rather than
an expert user), is a somewhat mature technique within steady state simulations
(including RANS); for LES, however, automatic, algorithm-driven
grid-adaptation is a new area of research.

While grid-adaptation for RANS is solely aimed at reducing the
numerical errors, it reduces both the numerical and subgrid modeling
errors in LES as well as increases the fraction of resolved motions.
This differentiates the problem from existing steady-state approaches,
and poses new challenges.
Pope~\cite{pope:04} suggested adaptation targeted to resolving a
user-defined fraction of the kinetic energy.
The many existing unsteady adaptive mesh refinement (AMR) algorithms
are most
commonly used in moving-front problems (explosions, flames, shocks, \ldots)
for which the adaptation criteria are quite different from those of
broadband turbulence.

\begin{figure}[!t]
\begin{minipage}[b]{0.68\textwidth}
  \centering
  \includegraphics[width=\textwidth,clip=true,trim=0pt 1mm 0pt 23cm]{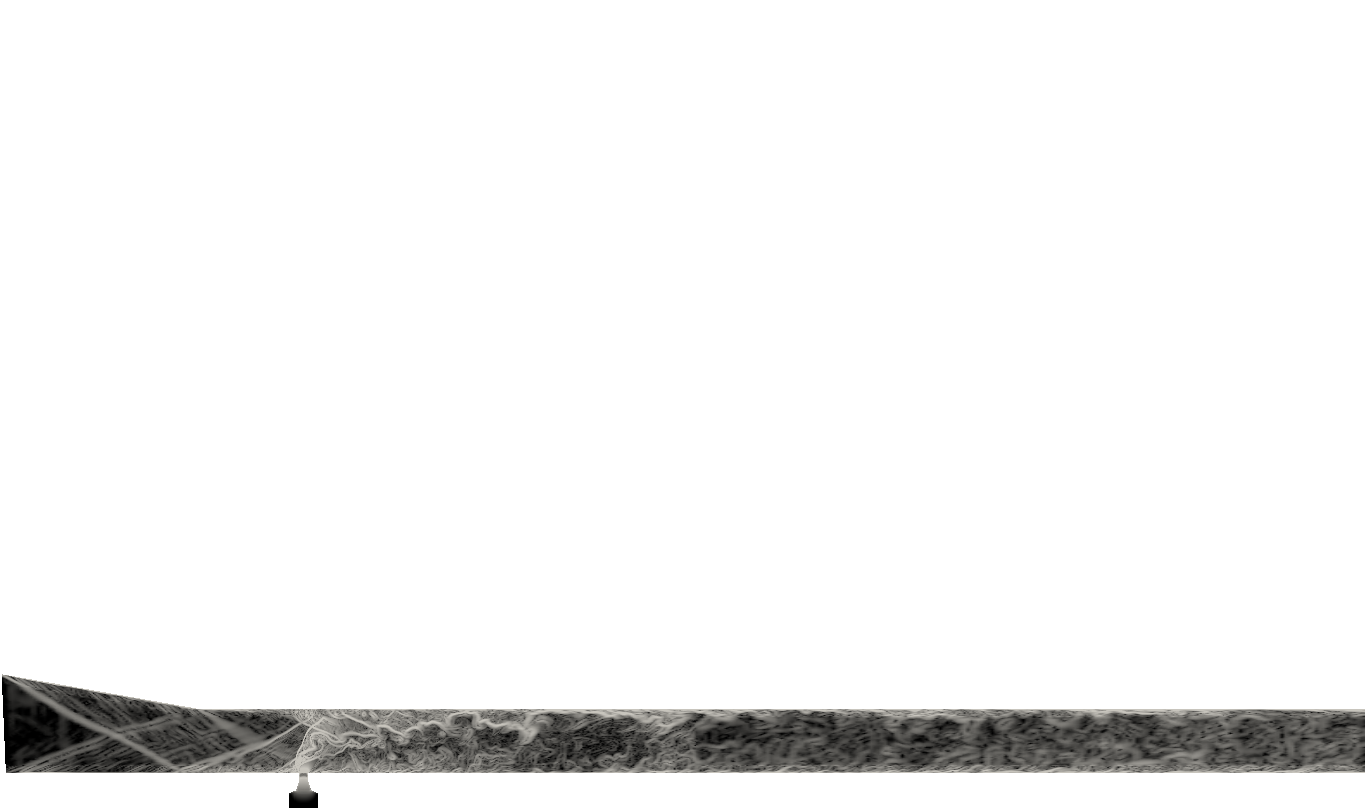}
  \begin{overpic}[width=\textwidth,clip=true,trim=0pt 0pt 0pt 167mm]{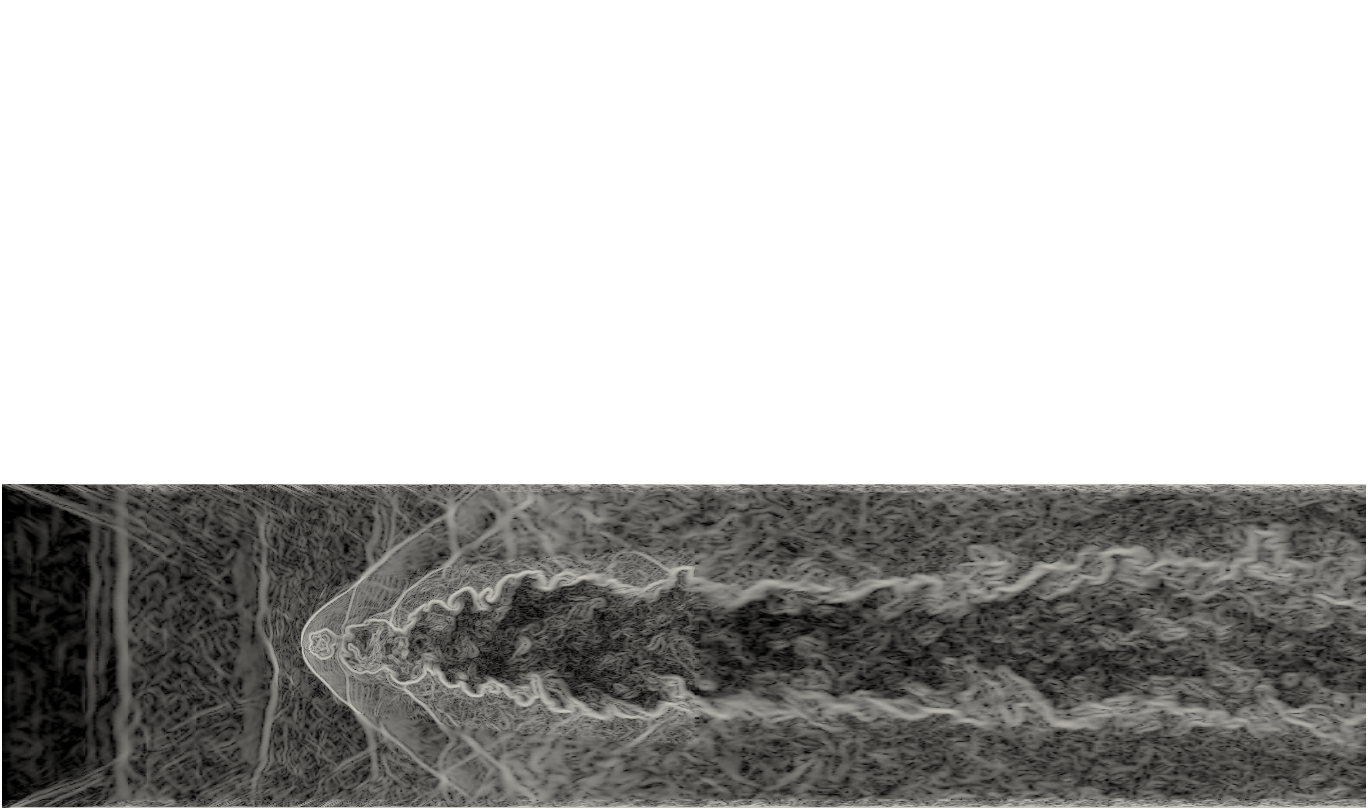}
    \linethickness{1.2pt}
    \put(36,22){\color{blue}
      \qbezier( 4, 0)( 4, 4)( 0, 4)
      \qbezier( 0, 4)(-4, 4)(-4, 0)
      \qbezier(-4, 0)(-4,-4)( 0,-4)
      \qbezier( 0,-4)( 4,-4)( 4, 0)
}
  \end{overpic}
\end{minipage}
\begin{minipage}[b]{0.31\textwidth}
  \centering
  \includegraphics[width=\textwidth,clip=true,trim=20mm 9mm 13mm 1mm]{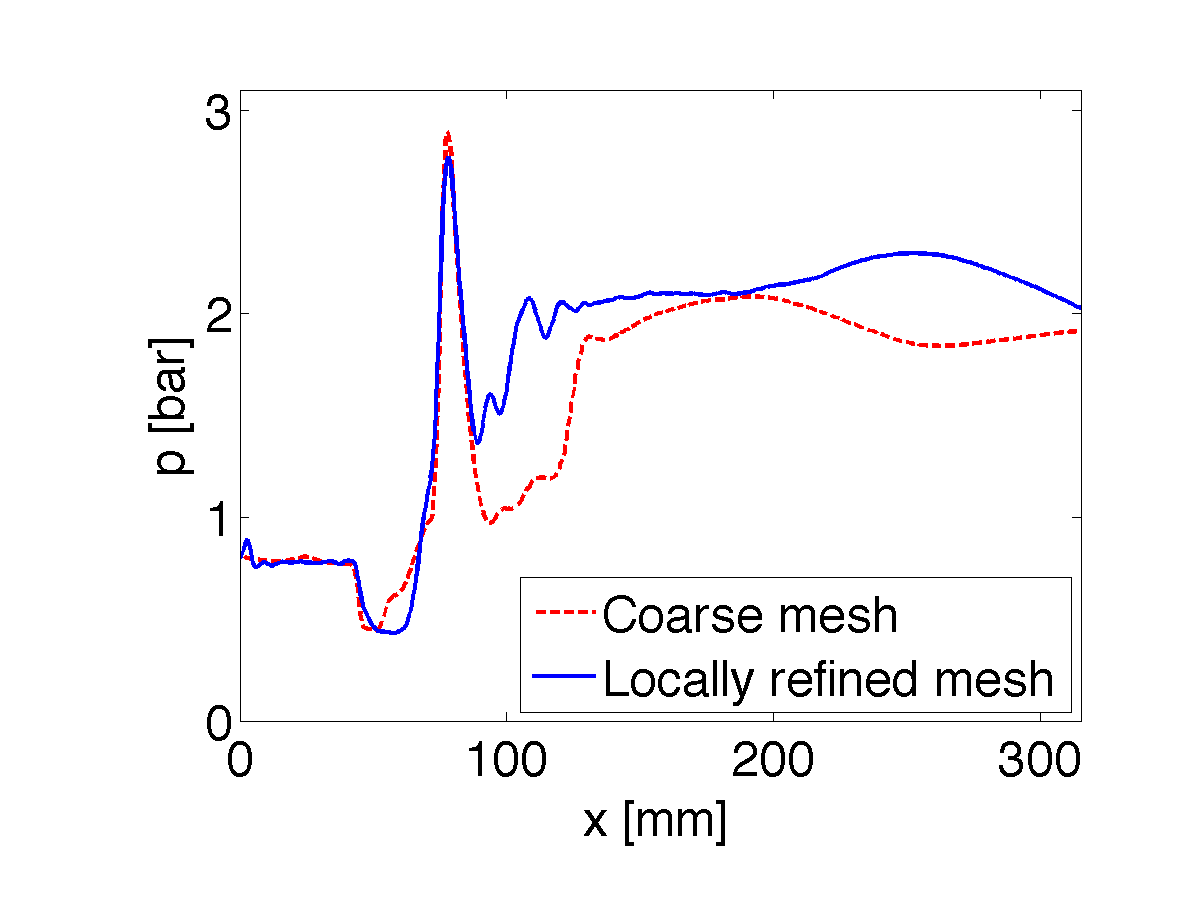}
\end{minipage}
\caption{ \label{fig:cesco}
  Illustrating the difficulty of generating grids that resolve the
  right features in the right places.
  The model scramjet combustor by Gamba \emph{et al.}~\cite{gamba:12}, with
  air at Mach 2.8 and 1100 K entering from the left and mixing with
  hydrogen injected from the bottom.
  The shocks and fuel/oxidizer mixing layers are seen in the simulated
  Schlieren images.
  The circular shape shows the main shock/boundary-layer interaction on the
  side wall, to be discussed in the text.
  To the right is shown the average pressure along the top wall from
  LES on two grids with 6M and 48M cells, respectively.
  The refinement was done in a manual, user-defined way, guided solely
  by experience and trial-and-error; the consequence of this is
  discussed in the text.
}
\end{figure}

To illustrate both the importance and difficulty of the
grid-generation process in LES, consider Fig.~\ref{fig:cesco}
which shows some sample results from an attempt
to perform LES of the reacting flow in a scramjet combustor.
The ``coarse'' grid in the figure was the initial grid.
Local grid-refinement (completely user-driven) to better resolve
different flow features was then attempted; this was initially focused
on the fuel injection and mixing region, but after much
trial-and-error it became clear that the grid resolution
in the region where the bow shock interacts with the side wall
boundary layer
(indicated by the circular shape in the figure)
was actually equally important
--
without refinement there, the shocks reflected back along the
centerline in the wrong place, thus causing delayed pressure rise
(from $x\approx 100$ mm to $x\approx 120$ mm).
It is difficult even for experienced users to anticipate exactly what
grid resolution is required in different regions, and thus a more
mathematical approach to this problem would quite clearly be very useful.

\subsection{Sensitivity analysis of chaotic flow simulations}

Sensitivity information, most obviously the gradient of the
quantity-of-interest with respect to all design variables, is
essential to optimization algorithms, uncertainty estimates and
analyses of how robust a design or prediction is.
The finite difference method to obtain sensitivites is popular and
easy to implement, but is surprisingly problematic in LES and DES.
This method approximates
derivatives by selecting a pair of adjacent design points,
then evaluating the quantity-of-interest at these points, and dividing
their difference by the distance between
the design points.
Most quantities-of-interest are mean quantities, and computed (i.e.,
sampled) averages converge to the true mean at a rate of $T_{\rm avg}^{-1/2}$
due to the central limit theorem of statistics.
When both objective functions are polluted by
sampling errors due to finite time-averaging,
the resulting
derivative is polluted by a large error, specifically twice the
sampling error divided by the distance between the design points.
Reducing this error sufficiently for accurate sensitivity information
could require impractically long simulations due to the
very slow
rate of convergence.

\begin{figure}[t!]\centering
  \subfigure[Instantaneous $J(t,s)$, computed as an initial value problem.]{
    \includegraphics[width=0.40\textwidth]{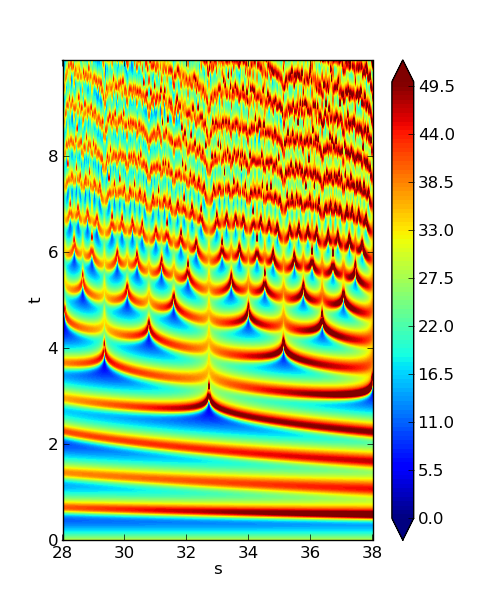}}
  \subfigure[Instantaneous $J(t,s)$, computed as a shadowing problem.]{
    \includegraphics[width=0.40\textwidth]{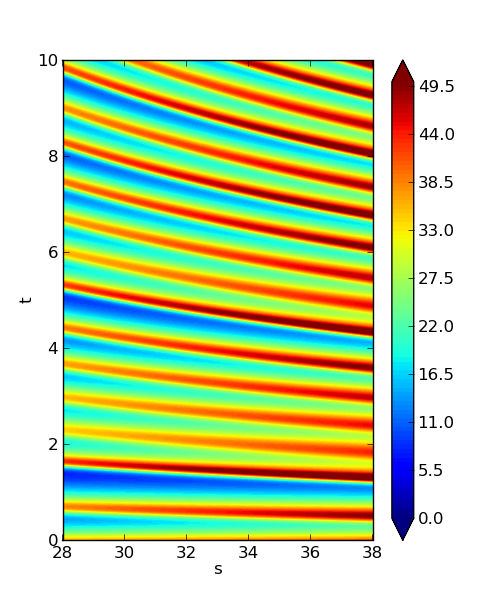}}
  \subfigure[Time-averaged $\overline{J}(s)$ of the initial value problem.]{
    \includegraphics[width=0.42\textwidth]{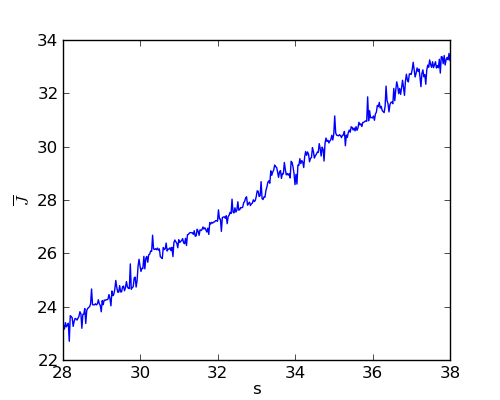}}
  \subfigure[Time-averaged $\overline{J}(s)$ of the shadowing problem.]{
    \includegraphics[width=0.42\textwidth]{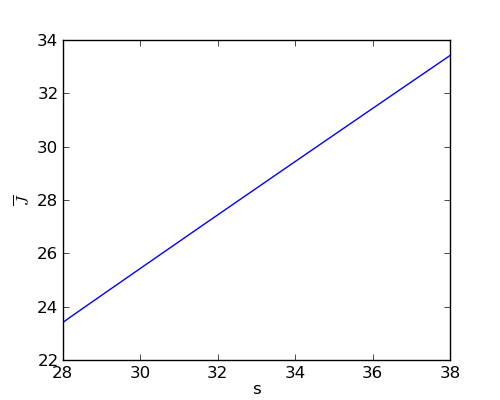}}
  \caption{
    Illustration of sampling errors when computing sensitivities in a
    chaotic system (the Lorenz system).
    The output functional $J$ as it varies with the design parameter $s$.
    The time-averages in the lower figures were computed over the
    full extent of time $t \in [0,10]$ in the upper figures.
    The ``noise'' in the left figures is the exponential divergence of nearby
    trajectories (slightly different values of $s$) -- the ``butterfly
    effect''.
  }
 \label{fig:lorenz_sensitivity}
\end{figure}

An example of this issue is shown in
Fig.~\ref{fig:lorenz_sensitivity}, which shows a specific
quantity-of-interest $J$ and how this varies in time and with the
design parameter $s$ for the Lorenz system
(for this problem, $J$ is simply the value of the third variable while $s$ is the
Rayleigh number).
When computed as an initial value problem, nearby trajectories
(starting from the same initial condition but having
slightly different values of $s$)
stay close for about 2-3 units of time and then rapidly become
completely uncorrelated.
When averaged, the uncorrelated sampling error results in a noisy
$\overline{J}(s)$ function which could not be differentiated in any
meaningful way.

In addition to the sampling error, the more standard problem with
finite difference sensitivities is that they require $N+1$ simulations
for $N$ design variables.
To overcome this limitation, Jameson~\cite{jameson:88}
introduced the adjoint method into CFD.
By simultaneously computing the derivative to many design variables,
the adjoint method has become widely used in aerodynamic design optimization.
In LES and DES, however, the adjoint method encounters a fundamental
challenge: the ``butterfly effect'' of chaotic dynamics.

By resolving the large scales of turbulence, LES inherits the chaotic dynamics
characteristic of turbulence.
A small perturbation, perhaps a slightly different mesh,
can utterly change the time history of a simulation.
This sensitivity to small purturbations (the ``butterfly effect'')
causes the adjoint method to diverge
when integrated over a long time period $T$.
The divergent adjoint solutions, observed
in a variety of flow solvers \cite{barthcylinder,wanggao,blonigan2012towards},
at first seem to suggest that perturbing the design always produces unbounded effects,
even on long-time averaged statistics.
This, of course, is not true: we know from experience that
a quantity-of-interest (properly: a statistic)
is, under the ergodic assumption,
sensitive only to the design parameters and not to the exact
realization (i.e., the initial condition).
The differentiability of ergodic statistics to a design parameter can
be proven under strict assumptions \cite{springerlink10}.

When computing the adjoint, however, the linearization inherent in the
definition of the adjoint implies that it is defined following the
specific solution trajectory along which it is computed.
The positive Lyapunov exponent(s) of chaotic systems
implies that there exists exponentially diverging trajectories;
this causes the adjoint to diverge as well.

\begin{figure}[t!]
  \centering
  \includegraphics[width=0.4\textwidth,clip=true,trim=0mm 6mm 0mm 0mm]{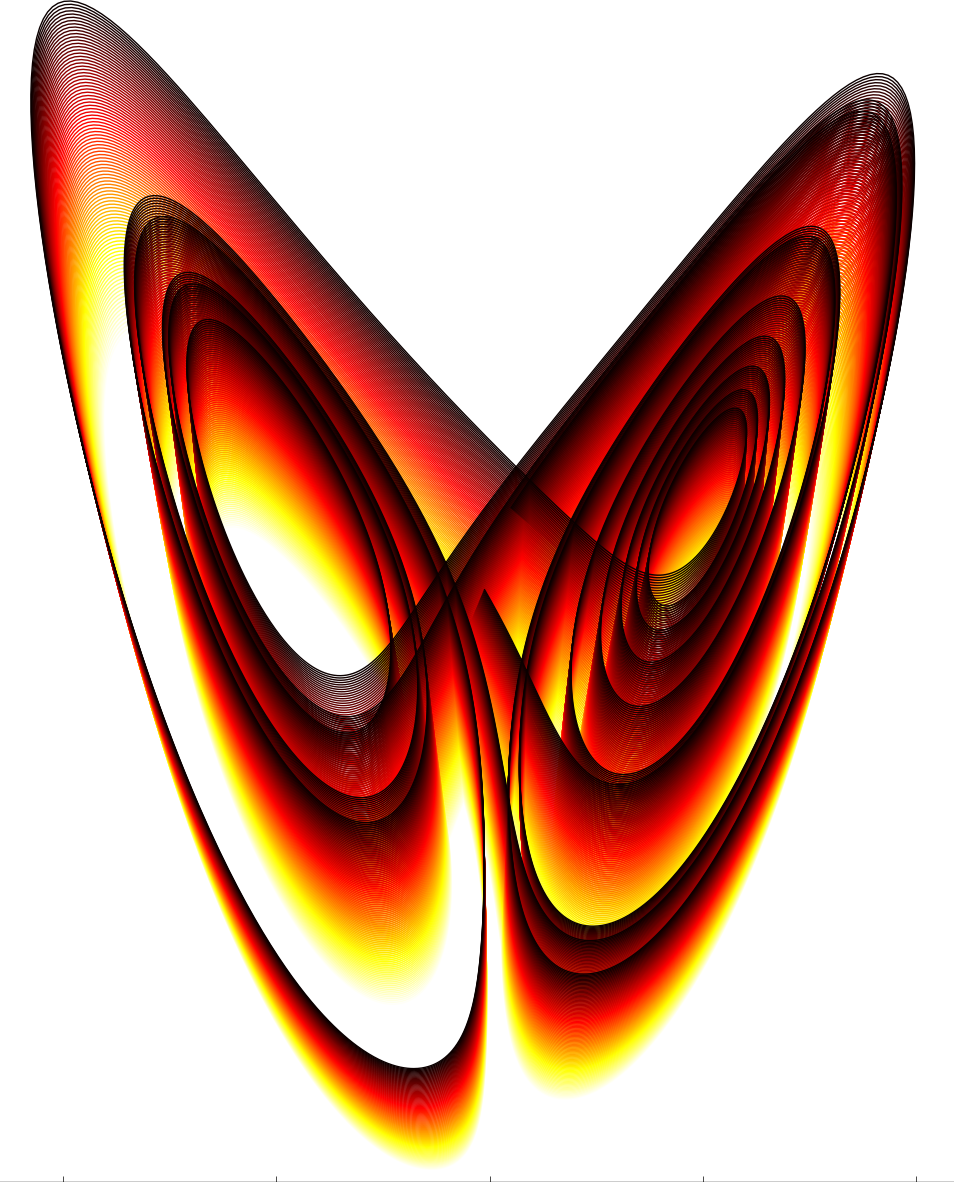}
  \caption{ \label{fig:shadows}
    Illustration of the shadowing paradigm, breaking causality:
    different trajectories in the Lorenz system at different values
    of the parameter $s$
    (shown with different colors).
    Note that the output functional $J$ in
    Fig.~\ref{fig:lorenz_sensitivity} is simply the average vertical
    position in this plot, thus clearly showing a variation in $J$
    for different $s$.
  }
\end{figure}

To overcome this problem, we must decouple the
sensitivity of statistics from the realization of the flow field.
A promising approach to this question is the shadowing
approach~\cite{lsstheory}.
The idea is to linearize the governing equation of an LES or DES, but
not the initial condition.
Liberated from the initial condition, the
resulting linearized equation has many
solutions.  They describe how different the flow over
a fixed pair of similar designs can be.
Almost all these solutions diverge exponentially, but some
of them do not.
The ones that stay bounded are called \emph{shadowing solutions}
-- they follow or ``shadow'' a given trajectory (see Fig.~\ref{fig:shadows}).
As the length of the simulation increases, not only does a
time averaged quantity converge to the infinite time average, but also
its computed derivative converges to the derivative of the infinite time
average \cite{lsstheory}.

The Least Square Shadowing method and its adjoint has been applied to a
range of dynamical systems, the largest being an isotropic
homogeneous turbulent flow simulation
at Taylor microscale Reynolds
number $Re_\lambda=33$ by a $32^3$
Fourier pseudo-spectral discretization~\cite{gomezthesis}.

Without useful sensitivities of long-time averaged quantities,
engineers are forced to drastically reduce the number
of design variables, therefore significantly reducing the utility of LES.
How to compute useful gradients in design-variable space, so that gradient-based optimization can
be applied to LES and DES, is a key research question.

In addition to enabling gradient-based optimization, sensitivities
would constitute an
important component in a more rigorous uncertainty estimation and
solution verification framework.
This is also true for grid-adaptation methods.

\subsection{Space-time parallelism}

The growth of computing power is projected to continue primarily
through increasing parallelism, as has been the case for more than a
decade now.
The CFD community has primarily
harnessed this growth through use of spatial domain decomposition,
but this strategy will become insufficient at some point.
Most algorithms are efficiently parallelized only for sufficiently
many cells per compute-process (core, thread, \ldots)
$n=N/p$, where $N$ and $p$ are the total number of cells and
compute-processes, respectively.
Thus there is a lower bound on the total wall-clock time that is
proportional to the number of time steps $N_t$,
which typically grows with $N$
for stability and accuracy reasons.

This may seem a distant problem, but it is arguably already becoming
an issue for certain problems.
For example, 
Pirozzoli and Bernardini (personal communication and~\cite{pirozzoli:11})
used 11M cells and needed 2M time steps to
capture the low-frequency oscillations in a shock/boundary-layer
interaction problem.
If the minimum useable $n$ is 1000, and a rather typical 50 $\mu$s per
cell and time step is required by the solver, then this calculation cannot be
completed in less than 28 hours on any machine available today --
despite the modest grid size.
And increasing the resolution would make it correspondingly worse, at
a rate of about $\sim N^{1/3}$.
Assuming continued growth in our appetite for ever larger grids,
there is an upper bound to the benefits possible from extreme
spatial parallelization.
One possible answer is to improve the floating-point utilization of
our codes (section~\ref{sec:hardware});
another is to parallelize the simulation in time as well as in space.

\begin{figure}[t!]
  \centering
  \includegraphics[width=0.7\textwidth]{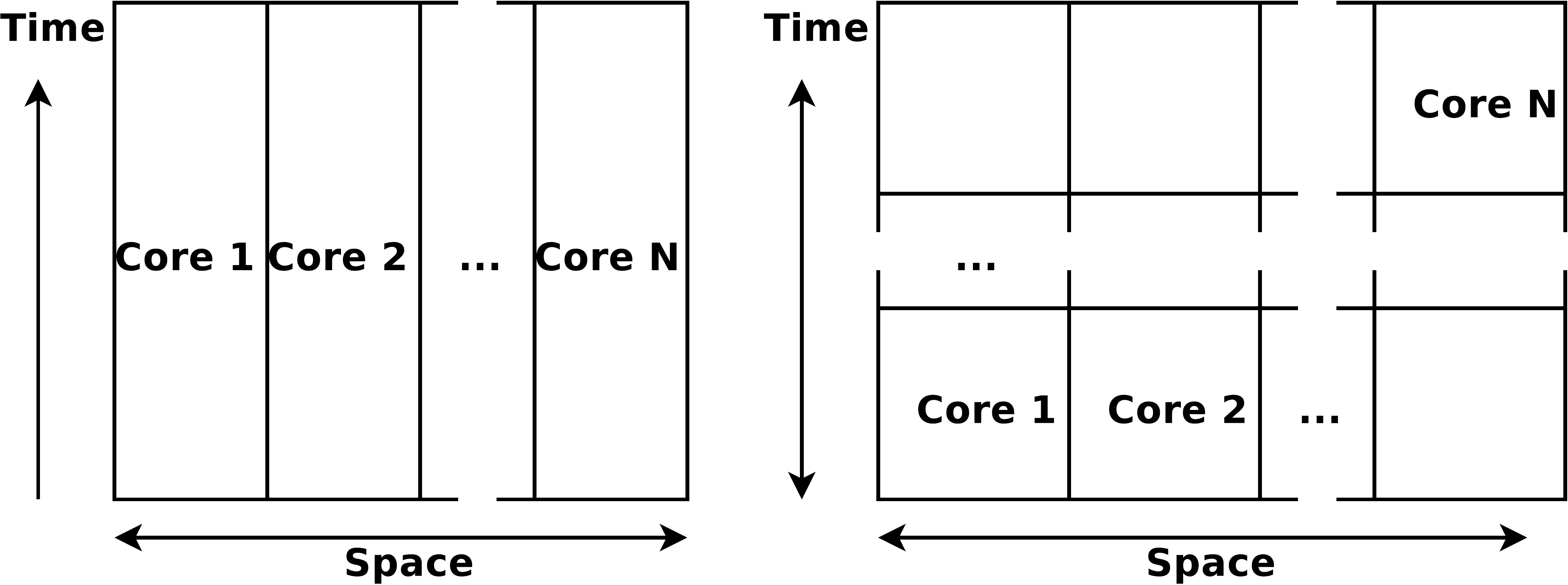}
  \caption{
    \label{f:spacetime}
    Illustration of spatial (left) and space-time (right)
    parallelism.}
\end{figure}

A space-time parallel simulation partitions the 4D
space-time, with each resulting subdomain assigned to a computing
process as illustrated in Fig.~\ref{f:spacetime}.
Time domain decomposition methods have a long history
\cite{nievergelt1964parallel}.
They include the multiple shooting method, the time-parallel and
space-time multigrid methods, and the Parareal method.
These methods start by computing an initial estimate of the solution
(in space and time) using a cheap, reduced-order solver.
This is then refined iteratively until it converges to the solution of
the initial value problem.
The key point is that this is done in parallel, i.e., different chunks
of time are iterated upon simultaneously.

This approach becomes problematic, however, for chaotic initial value
problems.
The ``butterfly effect''
implies that the late-time solution is (instantaneously) very
sensitive to the solution at early times.
It would therefore be futile to even start
iterating at the very late times before having reached a converged solution
at early times; this, of course, destroys the parallel efficiency of
the method.
Although some turbulent initial value problems (plasma turbulence) have been successfully parallelized
by the Parareal method~\cite{Samaddar:2010:PTN:1831747.1831799,reynolds2012mechanisms}, 
there is clearly an upper bound.


To achieve efficient time domain parallelism for chaotic problems, we
need to forsake the initial value problem and break causality.
Unlike current simulations in which the past completely
determines the future, a causality-breaking simulation allows for the future
to determine the past -- it modifies the past to resolve conflicts in the
future.
It computes the
solution in a way not unlike how this paper was written:
first a rough initial draft (cheap initial solver),
then
iterative refinements to all sections (solutions on finer grids)
coupled with modifications to early sections
to ensure consistency in the later sections (modifying the early
solution to ensure that the governing equations are satisfied at later
times).

\begin{figure}[t!] \centering
    \includegraphics[width=0.49\textwidth]{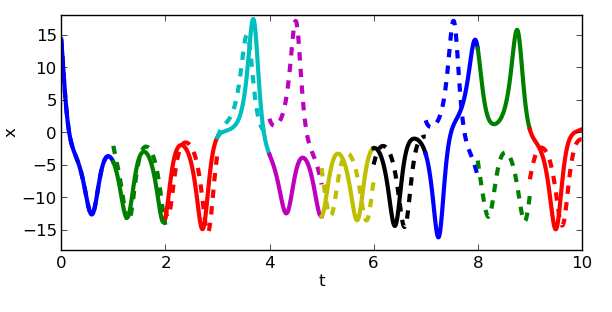}
    \includegraphics[width=0.49\textwidth]{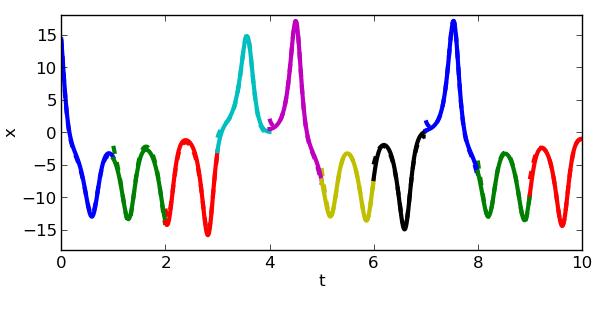}
\caption{ \label{f:ivplsp}
     Two different solutions to the Lorenz system (solid lines)
     starting from the same
     initial guess (dashed lines).
     The left plot enforces causality while
     the right one breaks it.
     Each color represents a time chunk that can be handled by a
     different computing core. The initial guess is inconsistent at the
     time chunk boundaries, and the converged solution must resolve all
     these inconsistencies. Note that the causality-enforcing simulation
     must make global, large changes to resolve small inconsistencies,
     while the causality-breaking simulation only need to make small,
     local changes.
}
\end{figure}

Mathematically, the causality-breaking solution is defined as the
solution closest (in the $L_2$-norm) to the initial guess that exactly
satisfies the governing equation.
The process is illustrated in Fig.~\ref{f:ivplsp}.
The initial guess is obtained by a reduced-order model and therefore
does not satisfy the governing equation.
Both the causality-enforcing and -breaking approaches reach final
solutions that satisfy the governing equation, but do so very
differently:
in the former, the small changes required at early times propagate
through the butterfly effect to produce very large changes at the
late times.
In contrast, only small changes are required in the causality-breaking
case, and (crucially) these changes are essentially local-in-time.

If this causality-breaking approach to time-parallelization could be
made to work on real LES problems, then it would make LES truly
scalable for ever larger problems and supercomputers.
More importantly, it would enable the solution of moderate-size (i.e.,
engineering design)
problems almost arbitrarily fast using extreme parallelization.

\subsection{Putting new LES technologies into engineering practice}

The preceeding sections listed some of the technical issues and
research areas the authors feel are particularly important and/or
promising in order to bring LES into practical use.
In addition, we must become better
at transitioning technologies from Academia to
corporate research labs and eventually to engineering design.

The LES community has not been as successful at this as
several other communities.
For example, the SST, SA and V2F turbulence models are all about 20 years
old, but have penetrated into most commercial CFD packages
and many engineering design processes.
The FFTW and Lapack libraries and the ParaView and VisIt visualization toolkits
were also developed within the last 20
years or so; they have become important enabling technologies used in a wide range of
fields.
In that same time frame, the LES community developed subgrid models, wall-models,
non-dissipative numerics, turbulent inflows and many other things --
and yet the impact of LES on engineering design has been much smaller.
So why have LES technologies been adopted more slowly, and what can be
done to speed things up?

While academic researchers should never become applied engineers, we
can and should make sure that the techniques we develop are applicable
to real engineering problems.
Novel modeling techniques should be fundamentally sound, and should be
thoroughly tested on canonical problems where detailed assessment is possible,
\emph{but} we should keep the final application in mind.
For example, if a subgrid model needs homogeneous directions for filtering
and/or averaging, then its use in most realistic geometries becomes
impossible.

The transitioning of knowledge should also include the transitioning
of enabling technologies, particularly computer code.
When the tax-paying public pays for a research project, it seems fair
that any code generated as part of it becomes publically available;
more to the point, this would enable further research at other
institutions, which surely should be our goal.
Progress on this front will come down to whether our
funding agencies and program managers start demanding it.

Ultimately, the LES community, academic and industrial, would benefit
from having a few well-documented, modular and capable open-source
codes.
OpenFOAM is a great example, but there is room for additional codes using
different modularization paradigms and languages.
Witness the impact of Matlab, which democratized matrix manipulations,
and imagine how CFD and LES research across the world could be elevated by
eliminating the constant drudgery of re-inventing the basic unstructured
CFD code.

\section{Final words}
When considering the role of LES in engineering design in 30 years,
one might wonder whether we will be able to replace RANS with LES
throughout the design process?
This is the wrong question to ask, and the answer is ``no'' anyway.
Better questions are:
``how and where in the process should we utilize LES to enable better
decision-making?''
and
``for which sub-systems or sub-domains would LES substantially increase our trust in
the predictions?''.
Phrased this way, three ``answers'' or observations emerge quite
clearly to the present authors:
1) there will definitely be a role for LES;
2) further advancement of the LES technology is needed to enable these
kinds of studies (e.g., technology enabling LES in a sub-domain with
accurate RANS $\leftrightarrow$ LES coupling);
and
3) beyond technology advancements, we need to show success stories
where LES clearly impacted design decision-making.

Several challenges multiply to make the use of LES in the
engineering design process difficult.
This is actually good news -- it means that
progress on these fronts, when combined, can have a large effect
on the feasibility of using LES in design optimization.
If improved time-stepping reduces the computational cost by a factor
of 10,
time-parallelization makes it possible to efficiently utilize 10 times
more cores,
and multi-fidelity optimization frameworks reduce the required number
of LES runs by another factor of 10,
then engineering LES will become 1000 times faster.
A good grid-adaptation framework would drastically reduce the
human-time required to build grids and simultaneously increase the
quality of the computed results.
Wall-models enable computations at realistic Reynolds numbers, and
methods to compute sensitivities enable optimization and uncertainty
estimation.
And surely there are other areas in which improvements can be made.
We are optimistic.

\section*{Acknowledgments}
JL is supported by a grant from the Minta Martin Foundation.
QW is supported by AFOSR Award F11B-T06-0007
under Dr. Fariba Fahroo, NASA Award NNH11ZEA001N under Dr. Harold
Atkins, a DOE ASCR Data-Centric Science Award under Sandy
Landsberg, and a GE contract with Dr. Sriram Shankaran and Dr.
Gregory M. Laskowski.

\bibliographystyle{elsart-num}
\bibliography{references}

\end{document}